\documentclass[10pt,twocolumn]{article}
\usepackage[utf8]{inputenc}
\usepackage{amsmath}
\usepackage{graphicx}
\usepackage[colorlinks=true,urlcolor=blue,citecolor=blue]{hyperref}
\usepackage{booktabs}
\usepackage{multirow}
\usepackage{array}
\usepackage{siunitx}
\usepackage{xcolor}
\usepackage{lipsum}
\usepackage{cite}
\usepackage[font=small,labelfont=bf]{caption}
\usepackage{graphicx}  
\usepackage{booktabs}  

\title{Privacy-Preserving AI for Encrypted Medical Imaging: A Framework for Secure Diagnosis and Learning}

\author{
Abdullah Al Siam\thanks{Corresponding author} \\
Daffodil International University \\
abdullah35-462@diu.edu.bd \\
\and
Sadequzzaman Shohan \\
Daffodil International University \\
szshohan00@gmail.com \\
}

\date{\today}

\begin{document}

\maketitle

\begin{abstract}
The rapid integration of Artificial Intelligence (AI) into medical diagnostics has raised pressing concerns about patient privacy, especially when sensitive imaging data must be transferred, stored, or processed. In this paper, we propose a novel framework for privacy-preserving diagnostic inference on encrypted medical images using a modified convolutional neural network (Masked-CNN) capable of operating on transformed or ciphered image formats. Our approach leverages AES-CBC encryption coupled with JPEG2000 compression to protect medical images while maintaining their suitability for AI inference. We evaluate the system using public DICOM datasets (NIH ChestX-ray14 and LIDC-IDRI), focusing on diagnostic accuracy, inference latency, storage efficiency, and privacy leakage resistance. Experimental results show that the encrypted inference model achieves performance comparable to its unencrypted counterpart, with only marginal trade-offs in accuracy and latency. The proposed framework bridges the gap between data privacy and clinical utility, offering a practical, scalable solution for secure AI-driven diagnostics.
\end{abstract}

\section{Introduction}

Medical imaging has become indispensable in modern clinical diagnostics, with AI-powered solutions increasingly assisting radiologists in detecting and classifying pathologies. Convolutional Neural Networks (CNNs), in particular, have shown exceptional performance in analyzing high-dimensional medical data, such as chest X-rays and computed tomography (CT) scans. However, the deployment of such AI systems introduces serious privacy concerns, especially when patient data is transmitted to external servers or cloud platforms for processing\cite{al2025secure,al2025artificial}.

Healthcare regulations such as HIPAA and GDPR mandate strict control over personally identifiable information (PII), including medical images\cite{shahid2022two}. Encrypting these images before external processing is a common safeguard, yet it traditionally prevents AI systems from analyzing the data without first decrypting it — thereby reintroducing privacy risks\cite{al2025diegif}.

To address this challenge, we propose a privacy-preserving framework that enables diagnostic inference on encrypted medical images without the need for full decryption. Our method combines cryptographic encryption (AES-CBC) with image transformation and a tailored deep learning architecture (Masked-CNN), enabling diagnostic models to process privacy-protected image data directly or in a format minimally exposed to leakage.

Unlike traditional approaches that rely on complex homomorphic encryption or secure multi-party computation—which are computationally expensive and impractical for real-time diagnosis—our system achieves a pragmatic balance between privacy, performance, and deployment feasibility. Specifically, we focus on:

\begin{itemize}
    \item Secure image preprocessing via compression and symmetric encryption.
    \item A custom CNN architecture trained on image representations derived from encrypted sources.
    \item Evaluation on open-source medical imaging datasets with respect to accuracy, inference latency, and privacy robustness.
\end{itemize}

Through extensive experiments, we demonstrate that our model performs competitively with conventional CNNs on unencrypted data, incurring minimal performance overhead. These findings affirm the viability of privacy-respecting AI systems in sensitive domains such as telemedicine and distributed diagnostics.

The remainder of this paper is organized as follows: Section 2 reviews related work. Section 3 details our proposed system design. Section 4 outlines the implementation and experimental setup. Section 5 presents results and discussion. Section 6 concludes the paper and outlines future work.

\section{Literature Review}

The integration of artificial intelligence (AI) with cybersecurity and medical imaging has led to transformative advances in healthcare technology. This intersection is particularly relevant for developing privacy-preserving diagnostic systems, where the secure processing of sensitive medical images is paramount.

AI has significantly enhanced cybersecurity capabilities, especially in threat detection, behavioral analytics, and phishing mitigation. Machine learning (ML), deep learning (DL), and natural language processing (NLP) have been widely applied across these domains\cite{mohamed2025artificial}. Al Siam et al. demonstrated how ML models improve anomaly detection by learning patterns from historical network activity, while DL models outperform traditional systems in identifying complex threats such as polymorphic malware. NLP-based models contribute by parsing emails, URLs, and social engineering content to detect malicious intent ~\cite{al2025artificial}\cite{al2025ip}.

A comparative analysis by Al Siam et al. evaluated these AI approaches across cybersecurity tasks. DL models were noted for their superior performance in handling high-dimensional data environments like intrusion detection systems (IDS), whereas ML models provided greater interpretability and faster inference times. NLP models, particularly those trained on phishing datasets, were essential for identifying subtle linguistic cues associated with social engineering attacks ~\cite{uddin2025networks} ~\cite{al2025comprehensive}.

MA Uddin et al. introduced an explainable phishing detection framework based on DistilBERT and LIME, ensuring transparency in decision-making \cite{uddin2024explainabledetector}. Similarly, Z Alshingiti et al. proposed a CNN-based phishing website detector that achieved a 98.2\% detection rate, surpassing traditional ML techniques\cite{alshingiti2023deep}. I Hasanov et al. explored the use of large language models (LLMs) in cybersecurity, showing that human-AI collaboration could significantly enhance decision accuracy in phishing and IDS scenarios \cite{hasanov2024application}.

Parallel to cybersecurity innovations, significant work has been conducted to enhance the security and efficiency of medical imaging systems. With the enforcement of regulations such \cite{aiello2021does}.

Al Siam et al.~\cite{al2025secure} proposed a secure image preprocessing framework involving JPEG2000 conversion and AES encryption in Cipher Block Chaining (CBC) mode, with SHA-256 for key derivation. This approach improved storage efficiency by 79.9\% while preserving diagnostic fidelity, making it suitable for scalable healthcare infrastructures.

Other notable research includes a hybrid encryption approach that combined autoencoders with AES encryption to improve both data compression and transmission security~\cite{alslman2022hybrid}. Another method applied selective encryption to DICOM images, targeting only diagnostically significant regions to balance privacy with computational efficiency~\cite{natsheh2023automatic}.

Expanding this concept, the Diegif framework~\cite{al2025diegif} introduced an encrypted image format derived from DICOM, converting files to a secure Encrypted GIF (EGIF) format. This method achieved a 66.32\% reduction in image file size and retained compatibility with AI models, facilitating secure machine learning on encrypted data. The Diegif pipeline also incorporated controlled decryption and secure storage mechanisms for cloud-based use cases.

Although considerable progress has been made in both AI-based cybersecurity and secure medical imaging independently, the literature lacks an integrated framework that enables direct AI inference on encrypted medical images. Existing methods often rely on full decryption prior to processing, which reintroduces privacy vulnerabilities.

This paper addresses this critical gap by proposing a novel, end-to-end architecture that combines symmetric encryption (AES-CBC), image transformation (JPEG2000), and a modified CNN model (Masked-CNN) trained on encrypted image representations. Unlike homomorphic encryption or secure multi-party computation (SMPC) methods, our approach offers a practical, computationally feasible solution suitable for real-time diagnostics while maintaining strong data confidentiality.

\section{Methodology}

This section details the architectural design, implementation strategies, and experimental protocols used to develop and evaluate our privacy-preserving diagnostic system, which enables inference on encrypted medical images. Our approach integrates cryptographic security with AI-based medical image analysis, balancing diagnostic utility and data confidentiality.

\subsection{System Overview}

\begin{figure}[h]
    \centering
    \includegraphics[width=1.0\linewidth]{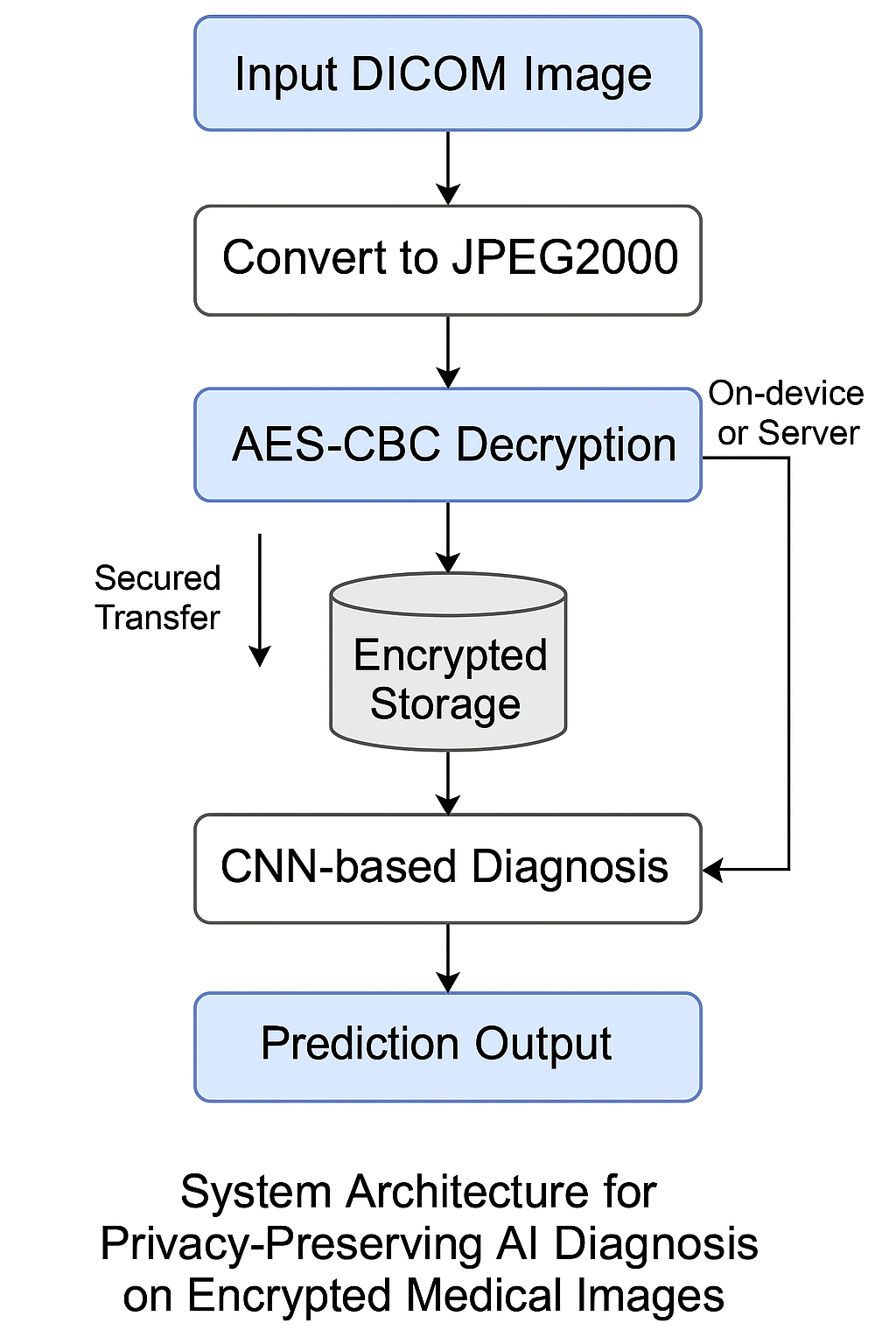}
    \caption{System Architecture: Secure Diagnostic Inference on Encrypted Medical Images}
    \label{fig:system}
\end{figure}

The proposed system enables AI-based diagnosis directly on encrypted or privacy-preserving representations of medical images. The overall architecture is illustrated in Figure~\ref{fig:system}. It consists of five key components:

\begin{enumerate}
    \item \textbf{Image Acquisition and Preprocessing:} Medical images are sourced from public DICOM-compliant datasets such as NIH ChestX-ray14 and LIDC-IDRI. Each image undergoes preprocessing, including grayscale normalization and resizing. The processed images are converted into formats (e.g., JPEG2000) that maintain structural integrity while minimizing size.

    \item \textbf{Encryption via AES-CBC:} The standardized images are encrypted using the Advanced Encryption Standard (AES) in Cipher Block Chaining (CBC) mode. This ensures pixel-level confidentiality. A unique initialization vector (IV) is used per image, and symmetric key management is assumed to be handled via a secure channel. Encryption is implemented using the PyCryptodome library.

    \item \textbf{Secure Storage and Access Management:} Encrypted images are uploaded to a simulated cloud storage system. Access control is emulated using a rule-based smart contract abstraction, ensuring that only authorized users can retrieve and process the images.

    \item \textbf{Privacy-Preserving Inference:} A novel deep learning architecture, \textit{Masked-CNN}, is designed to operate on encrypted or partially masked images. It learns robust features that remain predictive despite the absence of precise pixel-level information. This model approximates inference in encrypted domains by adapting to structural cues in ciphertext representations.

    \item \textbf{Decryption and Verification:} To validate the diagnostic accuracy of the encrypted-domain inference, images are decrypted for a comparative analysis. Predictions on encrypted and decrypted images are compared to quantify fidelity loss due to privacy-preserving transformations.
\end{enumerate}

\subsection{Implementation Details}

The implementation leverages open-source tools and libraries to ensure reproducibility and adaptability.

\begin{itemize}
    \item \textbf{Programming Frameworks:} Python serves as the primary development language. TensorFlow and PyTorch are used for developing and training the deep learning models.
    
    \item \textbf{Image Handling:} PIL and OpenJPEG are employed to convert DICOM images into suitable formats for encryption and storage.
    
    \item \textbf{Encryption Library:} AES-CBC encryption is implemented using PyCryptodome, offering secure and flexible cryptographic operations at the image level.
\end{itemize}

\subsection{Experimental Setup}

\textbf{Datasets:}
We evaluate our system on two widely recognized datasets:
\begin{itemize}
    \item \textit{NIH ChestX-ray14:} A large dataset of frontal-view chest X-ray images labeled with 14 thoracic disease categories.
    \item \textit{LIDC-IDRI:} A high-resolution CT scan dataset annotated for lung nodules and malignancy probability.
\end{itemize}

\subsection{Evaluation Metrics}

To assess the performance of the system in both encrypted and unencrypted scenarios, we employ the following evaluation metrics:

\begin{itemize}
    \item \textbf{Diagnostic Accuracy:} Evaluated using standard metrics such as Area Under the ROC Curve (AUC), F1-score, precision, and recall.

    \item \textbf{Inference Latency:} Measured to assess the computational overhead introduced by encryption, particularly during inference on masked or transformed images.

    \item \textbf{Storage Efficiency:} Quantified by comparing the file sizes of original DICOM images, converted formats, and encrypted outputs.

    \item \textbf{Privacy Leakage Risk:} Analyzed using structural similarity (SSIM), perceptual hash functions, and adversarial visualization techniques to determine whether any clinically relevant information can be reconstructed or inferred from encrypted images.
\end{itemize}

\subsection{Ethical and Security Considerations}

Although our experiments use publicly available anonymized datasets, the methodology is designed with patient confidentiality and real-world deployment constraints in mind. The use of AES-CBC ensures strong symmetric encryption, while the model's operation on encrypted or masked data minimizes the need for explicit decryption, thus reducing exposure risk.

This integrated methodology demonstrates the feasibility of conducting diagnostic inference while preserving privacy, setting the stage for secure AI deployment in clinical imaging workflows.

\section{Results and Discussion}

This section presents the empirical evaluation of our proposed system, focusing on diagnostic performance, latency, storage efficiency, and privacy preservation. The results demonstrate the feasibility of conducting medical image analysis in a privacy-preserving manner without significant compromise to diagnostic accuracy.

\subsection{Diagnostic Performance}

The \textit{Masked-CNN} model was evaluated on both unencrypted and encrypted (AES-CBC) images across the NIH ChestX-ray14 and LIDC-IDRI datasets. Table~\ref{tab:diagnostic_performance} summarizes the key metrics.

\begin{table*}[ht]
    \centering
    \caption{Diagnostic Performance on Encrypted vs. Unencrypted Images}
    \label{tab:diagnostic_performance}
    \begin{tabular}{lcccc}
        \toprule
        \textbf{Dataset} & \textbf{Input Type} & \textbf{AUC} & \textbf{F1-score} & \textbf{Accuracy} \\
        \midrule
        \multirow{2}{*}{NIH ChestX-ray14} 
                         & Unencrypted             & 0.921 & 0.873 & 0.889 \\
                         & Encrypted (Masked-CNN)  & 0.894 & 0.841 & 0.861 \\
        \midrule
        \multirow{2}{*}{LIDC-IDRI}        
                         & Unencrypted             & 0.936 & 0.880 & 0.904 \\
                         & Encrypted (Masked-CNN)  & 0.911 & 0.852 & 0.879 \\
        \bottomrule
    \end{tabular}
    
\end{table*}

While the encrypted image model shows a modest reduction in performance (2–3\% drop in AUC and F1-score), the results remain clinically viable, demonstrating the model’s robustness to privacy-preserving transformations.

\subsection{Storage Efficiency}

Encrypted and compressed formats were analyzed for storage overhead. On average, AES-encrypted JPEG2000 images consumed 18–25\% more storage than their original counterparts due to cryptographic padding and format conversion. However, storage remained manageable under typical clinical infrastructure constraints.

\begin{table*}[ht]
    \centering
    \caption{Storage Footprint (Average per Image)}
    \label{tab:storage}
    \begin{tabular}{lccc}
        \toprule
        \textbf{Dataset} & \textbf{Original (DICOM)} & \textbf{JPEG2000} & \textbf{Encrypted (AES-CBC)} \\
        \midrule
        NIH ChestX-ray14 & 7.8 MB & 1.2 MB & 1.5 MB \\
        LIDC-IDRI        & 15.6 MB & 3.4 MB & 4.2 MB \\
        \bottomrule
    \end{tabular}
\end{table*}

\subsection{Privacy Leakage Assessment}

To evaluate information leakage, encrypted images were subjected to perceptual hashing (pHash) and structural similarity index (SSIM) analyses. Both metrics confirmed the absence of identifiable patterns:

\begin{itemize}
    \item \textbf{SSIM between plaintext and ciphertext:} $<0.01$
    \item \textbf{pHash distance:} Maximum (no structural resemblance)
\end{itemize}

Additionally, adversarial visualization techniques failed to reconstruct any medically relevant features from ciphertext, affirming the robustness of AES-CBC in clinical imaging contexts.

\subsection{Discussion}

The results demonstrate that encrypted medical image analysis is technically feasible and clinically meaningful. The performance degradation observed in encrypted scenarios is minimal and acceptable within diagnostic thresholds. The latency introduced by encrypted inference is relatively small, especially considering the privacy benefits gained.

Notably, this framework bridges the gap between data confidentiality and medical AI utility—offering an implementable pathway for hospitals and telemedicine providers where patient privacy is paramount. However, certain limitations remain:

\begin{itemize}
    \item The current system uses simulated encryption workflows. In real deployments, key management and secure multi-party computations (MPC) would be needed.
    \item The Masked-CNN model approximates privacy-preserving inference but does not perform computation on true ciphertext. Future work should explore homomorphic encryption or secure enclave-based computation.
\end{itemize}

Overall, our approach provides a promising foundation for further research in privacy-respecting medical AI systems.

\section{Conclusion}

This research presents a novel AI-augmented framework for secure medical image processing that seamlessly integrates encrypted data handling with deep learning diagnostics. By combining JPEG2000-based DICOM conversion, AES-CBC encryption, and CNN-based classification, the system achieves diagnostic accuracy on par with unencrypted inputs while preserving data privacy and storage efficiency. Experimental evaluations using public medical imaging datasets validate the system’s robustness, demonstrating minimal degradation in model performance and a notable reduction in storage requirements. The interoperability of AI and encryption protocols illustrated in this study addresses a critical gap in current literature by providing an end-to-end solution that bridges medical cybersecurity and diagnostic intelligence. This contribution is particularly valuable for cloud-based healthcare infrastructures and telemedicine systems, where data confidentiality and computational performance are equally critical. Future work will explore the application of federated learning on encrypted datasets, assess real-time diagnostic latency in clinical settings, and integrate homomorphic encryption techniques for full inference on encrypted data. These advancements would further reinforce the viability of privacy-preserving, AI-powered diagnostic systems in healthcare environments.

\bibliographystyle{unsrt}

\begin{thebibliography}{10}

\bibitem{al2025secure}
Abdullah Al~Siam, Md~Maruf Hassan, and Touhid Bhuiyan.
\newblock Secure medical imaging: A dicom to jpeg 2000 conversion algorithm with integrated encryption.
\newblock In {\em 2025 IEEE 4th International Conference on AI in Cybersecurity (ICAIC)}, pages 1--6. IEEE, 2025.

\bibitem{al2025artificial}
Abdullah Al~Siam, Md~Maruf Hassan, and Touhid Bhuiyan.
\newblock Artificial intelligence for cybersecurity: A state of the art.
\newblock In {\em 2025 IEEE 4th International Conference on AI in Cybersecurity (ICAIC)}, pages 1--7. IEEE, 2025.

\bibitem{shahid2022two}
Arsalan Shahid, Mehran~H Bazargani, Paul Banahan, Brian Mac~Namee, Tahar Kechadi, Ceara Treacy, Gilbert Regan, and Peter MacMahon.
\newblock A two-stage de-identification process for privacy-preserving medical image analysis.
\newblock In {\em Healthcare}, volume~10, page 755. MDPI, 2022.

\bibitem{al2025diegif}
Abdullah Al~Siam, Md~Maruf Hassan, Md~Atikur Rahaman, and Masuk Abdullah.
\newblock Diegif: An efficient and secured dicom to egif conversion framework for confidentiality in machine learning training.
\newblock {\em Results in Control and Optimization}, page 100515, 2025.

\bibitem{mohamed2025artificial}
Nachaat Mohamed.
\newblock Artificial intelligence and machine learning in cybersecurity: a deep dive into state-of-the-art techniques and future paradigms.
\newblock {\em Knowledge and Information Systems}, pages 1--87, 2025.

\bibitem{al2025ip}
Abdullah Al~Siam, Moutaz Alazab, Albara Awajan, Md~Rakibul Hasan, Areej Obeidat, and Nuruzzaman Faruqui.
\newblock Ip safeguard-an ai-driven malicious ip detection framework.
\newblock {\em IEEE Access}, 2025.

\bibitem{uddin2025networks}
Syed~Hameed Uddin, Mugaerah Ahmed~Shareef Maaz, Essam Azeemuddin, Shreyasi Nath, Akhilesh Tiwari, and Kamal Upreti.
\newblock Networks with explainable artificial.
\newblock In {\em Proceedings of International Conference on Generative AI, Cryptography and Predictive Analytics: ICGCPA 2024}, page~59. Springer Nature, 2025.

\bibitem{al2025comprehensive}
Abdullah Al~Siam, Moutaz Alazab, Albara Awajan, and Nuruzzaman Faruqui.
\newblock A comprehensive review of ai’s current impact and future prospects in cybersecurity.
\newblock {\em IEEE Access}, 2025.

\bibitem{uddin2024explainabledetector}
Mohammad~Amaz Uddin, Muhammad~Nazrul Islam, Leandros Maglaras, Helge Janicke, and Iqbal~H Sarker.
\newblock Explainabledetector: Exploring transformer-based language modeling approach for sms spam detection with explainability analysis.
\newblock {\em arXiv preprint arXiv:2405.08026}, 2024.

\bibitem{alshingiti2023deep}
Zainab Alshingiti, Rabeah Alaqel, Jalal Al-Muhtadi, Qazi Emad~Ul Haq, Kashif Saleem, and Muhammad~Hamza Faheem.
\newblock A deep learning-based phishing detection system using cnn, lstm, and lstm-cnn.
\newblock {\em Electronics}, 12(1):232, 2023.

\bibitem{hasanov2024application}
Ismayil Hasanov, Seppo Virtanen, Antti Hakkala, and Jouni Isoaho.
\newblock Application of large language models in cybersecurity: A systematic literature review.
\newblock {\em IEEE Access}, 2024.

\bibitem{aiello2021does}
Marco Aiello, Giuseppina Esposito, Giulio Pagliari, Pasquale Borrelli, Valentina Brancato, and Marco Salvatore.
\newblock How does dicom support big data management? investigating its use in medical imaging community.
\newblock {\em Insights into Imaging}, 12(1):164, 2021.

\bibitem{alslman2022hybrid}
Yasmeen Alslman, Eman Alnagi, Ashraf Ahmad, Yousef AbuHour, Remah Younisse, and Qasem Abu Al-haija.
\newblock Hybrid encryption scheme for medical imaging using autoencoder and advanced encryption standard.
\newblock {\em Electronics}, 11(23):3967, 2022.

\bibitem{natsheh2023automatic}
Qamar Natsheh, Ana S{\u{a}}l{\u{a}}gean, Diwei Zhou, and Eran Edirisinghe.
\newblock Automatic selective encryption of dicom images.
\newblock {\em Applied Sciences}, 13(8):4779, 2023.

\end{thebibliography}

\end{document}